\documentclass[11pt,twoside]{article}

\usepackage{a4}
\usepackage[utf8]{inputenc}
\usepackage{amssymb,amsfonts,graphicx,color,subfig,ae}
\usepackage{amsthm}
\usepackage[english]{babel}
\usepackage[intlimits]{amsmath}
\usepackage{eucal}
\usepackage{cite}
\usepackage{hyperref}
\usepackage[heightrounded]{geometry}

\newcommand{\2}{{\textstyle \frac{1}{2}}}
\newcommand{\mcH}{\mathcal{H}}

\newcommand{\ve}{\varepsilon}

\newcommand{\dx}[1]{\mathrm{d}#1}

\newcommand{\mfa}{\mathfrak{a}}

\newcommand{\mfbq}{\bar{\mathfrak{b}}}

\newcommand{\s}{\sigma}

\newcommand{\epp}{\, .}
\newcommand{\epc}{\, ,}

\def\beq{\begin{equation}}
\def\eeq{\end{equation}}
\def\bea{\begin{eqnarray}}
\def\eea{\end{eqnarray}}

\def\beann{\begin{eqnarray*}}
\def\eeann{\end{eqnarray*}}

\let\a=\alpha

   \let\m=\mu
  \let\p=\pi \let\r=\rho \let\s=\sigma

  \let\D=\Delta

\let\qd=\quad  

\def\epp{\, .}
\def\epc{\, ,}

\theoremstyle{plain}

\newtheorem*{corollary*}{Corollary}

\newtheorem*{conjecture*}{Conjecture}

\theoremstyle{definition}

\def\2{\frac{1}{2}} \def\4{\frac{1}{4}}

\def\6{\partial}

\def\+{\dagger}

\def\<{\langle} \def\>{\rangle}

\def\i{{\rm i}}

\def\rd{{\rm d}}
\def\re{{\rm e}}

\def\ctg{\, {\rm ctg}\,}

\renewcommand{\cot}{\ctg}

\DeclareMathOperator{\sh}{sh}
\DeclareMathOperator{\ch}{ch}

\DeclareMathOperator{\tr}{tr}

\def\diag{{\rm diag}}

\title{Thermodynamics and short-range correlations of the XXZ chain
close to its triple point}

\author{
Christian Trippe\thanks{e-mail:
\href{mailto:trippe@physik.uni-wuppertal.de}%
{\protect\nolinkurl{trippe@physik.uni-wuppertal.de}}}\;,
Frank G\"ohmann\thanks{e-mail:
\href{mailto:goehmann@physik.uni-wuppertal.de}%
{\protect\nolinkurl{goehmann@physik.uni-wuppertal.de}}}\;,
Andreas Kl\"umper\thanks{e-mail:
\href{mailto:kluemper@physik.uni-wuppertal.de}%
{\protect\nolinkurl{kluemper@physik.uni-wuppertal.de}}}
\\
\parbox{0.9 \linewidth}{\vspace{0.4 \baselineskip}\centering
    Fachbereich C -- Physik, Bergische Universit\"at Wuppertal,\\
    42097 Wuppertal, Germany}
}

\date{\today}

\begin{document}

\maketitle

\begin{abstract}
The XXZ quantum spin chain has a triple point in its ground state
$h$-$1/\D$  phase diagram. This first order critical point is located
at the joint end point of the two second order phase transition
lines marking the transition from the gapless phase to the fully polarized
phase and to the N\'eel ordered phase, respectively. We explore the
magnetization and the short-range correlation functions in its vicinity
using the exact solution of the model. In the critical regime above the
triple point we observe a strong variation of all physical quantities on
a low energy scale of order $1/\D$ induced by the transversal quantum
fluctuations. We interpret this phenomenon starting from a strong-coupling
perturbation theory about the highly degenerate ground state of the Ising
chain at the triple point. From the perturbation theory we identify the
relevant scaling of the magnetic field and of the temperature. Applying
the scaling to the exact solutions we obtain explicit formulae for the
magnetization and short-range correlation functions at low temperatures.
\end{abstract}

\section{Introduction}
Critical points in the ground state phase diagram are nowadays often
called quantum critical. In this terminology the triple point of the
XXZ chain is a quantum critical point of first order. Quantum critical
points have been proposed as a paradigm in condensed matter physics for
explaining severe deviations in the behaviour of strongly correlated
electrons from the Fermi liquid picture. Much of the intuitive
understanding of quantum critical points comes from simple exactly
solved models \cite{Sachdev00}. In this work we would like to extend
the list of those quantum critical points that can be studied exactly
by another example.

The ground state phase diagram of the XXZ chain in the $h$-$1/\D$ plane
has three distinguished critical points. Two of them are the
isotropic ferromagnetic and antiferromagnetic points at $h = 0$ and
$\D = - 1$ or $\D = 1$, respectively. Although the model is exactly
solvable by Bethe ansatz, it is hard to calculate its thermodynamic
properties in the vicinity of the ferromagnetic point. The
antiferromagnetic point is easier to access, but still requires a certain 
effort in the numerical solution of the integral equations \cite{TGK09pp}.

A third critical point, to which less attention has been paid, is the
aforementioned triple point. It will be studied in this work. The
triple point is located at $\D = \infty$ and $h = h_c$ with $\D$ denoting
the anisotropy parameter of the XXZ model and $h_c$ the critical magnetic
field separating between the fully polarized phase and the N\'eel phase
of the Ising chain. This critical point is simpler than the other two,
because degenerate strong-coupling perturbation theory can be applied in
its vicinity. Applying the same perturbative expansion to the integral
equations that determine the thermodynamics of the model \cite{TGK09pp}
we obtain free fermion type equations for low temperatures and
algebraic equations in the zero temperature limit. Those equations provide
explicit results e.g.\ for the ground state magnetization and some
static correlation functions up to first order in $1/\D$. On the other
hand, treating the integral equations numerically, we obtain the exact
temperature dependent picture, and we can assess the range of validity
of the perturbation theory.

\section{XXZ chain in the vicinity of the strong coupling critical point}
\label{sec:hams}
We consider the XXZ Hamiltonian, which on the `gapped antiferromagnetic
side' of the phase diagram is most naturally written as
\begin{equation} \label{ham}
     \mcH=\sum_{j=1}^L
        \left[
        J(\sigma_{j-1}^z\sigma_{j}^z-1) + \frac{J}{\D}
	(\sigma_{j-1}^x\sigma_{j}^x+\sigma_{j-1}^y\sigma_{j}^y)
        - \frac{h}{2} \s_j^z
	\right]
\end{equation}
with $J, h >0$ and $\Delta>1$. Here the $\sigma_j^\alpha, \, j=1,\ldots,
L$, $\a = x, y, z$, act locally as Pauli matrices, $J$ is the exchange
coupling, $h$ denotes the strength of the magnetic field, and
$\Delta=\ch(\eta)$ is the anisotropy para\-meter.

\begin{figure}[t]
\begin{center}
\includegraphics{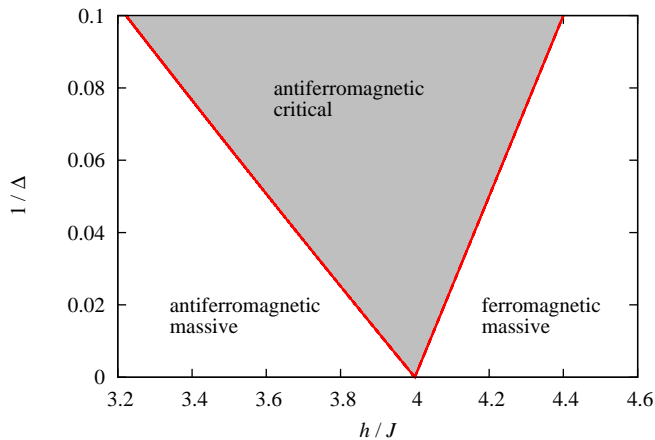}
\end{center}
\caption{(Color online) Ground state phase diagram of the XXZ chain
in the $h$-$1/\Delta$ plane for fixed $J$ and large $\Delta$ in the
vicinity of the triple point.}
\label{fig:phasendiagramm_ising}
\end{figure}
The ground state phase diagram in the $h$-$1/\Delta$ plane can be obtained
from the Bethe ansatz solution of the model (see e.g.\ \cite{Takahashi99}).
We show a detail of it in figure~\ref{fig:phasendiagramm_ising}. The
lower critical field $h_e$ (left red line) is given by the excitation gap
\cite{DeGa66,YaYa66d} whereas the saturation field (right red line) is
given by the simple explicit formula $h_s=4J(1+1/\Delta)$. These
fields determine the second order phase transition lines between the
critical and massive phases drawn in figure~\ref{fig:phasendiagramm_ising}.
The lines are asymptotically straight. For the lower critical field one
can see deviations from the asymptotic behaviour $h_e\approx4J(1-2/\Delta)$,
but they are small on the scale used in the figure. The point $h = 4J$,
$\D = \infty$ in which the two second order phase transition lines join
is the triple point we are interested in. In the following we shall
characterize it starting with the Hamiltonian.

For $\Delta\rightarrow\infty$ the XXZ chain (\ref{ham}) simplifies to
the Ising chain
\begin{equation} \label{hising}
     \mcH_I (h) = \sum_{j=1}^L\left[ J (\sigma_{j-1}^z\sigma_{j}^z-1)
        -\frac{h}{2} \sigma_j^z \right]
                = \2 \sum_{j=1}^L \diag (-h, -4J, -4J, h)_{j-1, j} \epc
\end{equation}
which is diagonal in a basis of tensor products of local $S^z$-eigenstates.
From the second equation we can easily understand how its ground states
change with the magnetic field. If $h < 4J$ the two N\'eel states
consisting of alternating up- and down-spins are the ground states.
If $h > 4J$ the ground state is the unique fully polarized state with
all spins pointing upwards. In the former case the ground state
magnetization per lattice site vanishes, and in the latter case it is one
half. Thus, there is a first order phase transition point at $h = h_c = 4J$.

What happens exactly at the phase transition point? As can be seen
from (\ref{hising}) all states with no neighbouring down-spins
are degenerate and have lowest energy $-2JL$. Clearly their number
is macroscopic in the thermodynamic limit where it is measured as the
average number $s$ of states per site. We can obtain it from the known
\cite{Babook} entropy per site in the zero temperature limit. It is equal
to the golden mean, $s = (1 + \sqrt{5})/2$. All states with exactly two
neighbouring down spins are the lowest excited states above the
ground state with excitation energy $4J$. The magnetization per lattice
site right at the critical point takes a value different from zero and
one half, namely $m = 1/2 \sqrt{5}$.

Let us rewrite the Hamiltonian \eqref{ham} as
\begin{equation} \label{eq:ham_rewritten}
     \mcH= \mcH_I (h_c) +\frac{J}{\Delta} \sum_{j=1}^L
        \left(\sigma_{j-1}^x\sigma_{j}^x+\sigma_{j-1}^y\sigma_{j}^y
        - 2 \a \s_j^z \right)
\end{equation}
with $\a = \D (h/h_c - 1)$. Then, for every fixed $\a$, the second term
becomes small for sufficiently large $\D$ and can be viewed as
a perturbation to the Hamiltonian $\mcH_I (h_c)$. The splitting
of the ground state under the influence of the perturbation can
be understood by means of an effective Hamiltonian obtained in second
order degenerate perturbation theory, in a similar way as e.g.\ the
Heisenberg Hamiltonian is obtained from the Hubbard Hamiltonian
at half band-filling in the strong coupling limit (see e.g.\ appendix 2.A
of \cite{thebook}). Note that curves of fixed $\a$ are straight lines
originating from the critical point at $h = h_c$ and $\D = \infty$ in
figure~\ref{fig:phasendiagramm_ising}.

Using second order degenerate perturbation theory we obtain the effective
Hamiltonian
\begin{multline} \label{efham}
     \mcH= -2JL + \frac{J}{\Delta} P_0 \sum_{j=1}^L
           \left(\sigma_{j-1}^x\sigma_{j}^x
	         +\sigma_{j-1}^y\sigma_{j}^y - 2 \alpha \s_j^z \right) P_0\\
	- \frac{J}{\Delta^2} P_0\sum_{j=1}^L
        \left[\frac{1}{2}\left(1-\sigma_{j-1}^z\right)
	      \left(1-\sigma_{j+1}^z\right)
	+\left(\sigma_{j-2}^+\sigma_{j-1}^-\sigma_j^+\sigma_{j+1}^-
	       +h.c.\right) \right]P_0 \epc
\end{multline}
describing the splitting of the ground state. Here $P_0$ is the projector
onto the subspace with no neighbouring down-spins. Note that the
magnetic field does not enter into the second order correction.

The first order term, let us call it $\mcH_1$, is an integrable Hamiltonian
\cite{AlBa99b} in full analogy with the situation with the one-dimensional
Hubbard model in the strong coupling limit. The operator $\r_1 =
\re^{-J \mcH_1/\D T}/\tr \re^{-J \mcH_1/\D T}$ replaces the statistical
operator to lowest order in $J/\D$. For $T \rightarrow 0+$ it becomes
the projector onto the ground state of $\mcH_1$. It follows that, to
leading order, the ground state correlation functions depend only on~$\a$,
i.e.\ they are constant on straight lines originating from the
triple point. Their values on lines parallel to the $h$-axis in
figure~\ref{fig:phasendiagramm_ising} must vary continuously between
the values taken at the boundaries between the massive and the massless
phases marked by the second order phase transition lines. The value of
the magnetization per lattice site, in particular, calculated in the limit
$\D \rightarrow \infty$ on straight lines of fixed $\a$, must vary
continuously between zero and one half, which already gives us a precise
picture of the nature of the singularity of the first order quantum
critical point.

\begin{figure}[t]
\begin{center}
\includegraphics{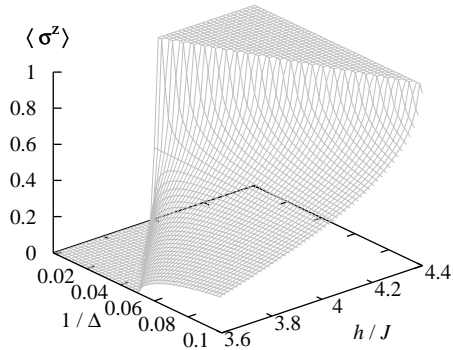}
\end{center}
\caption{Magnetization of the XXZ chain close to the triple point
for $T \rightarrow 0$.}
\label{fig:mag3d}
\end{figure}
Considering the finite temperature physics of the effective Hamiltonian
(\ref{efham}) for fixed~$\a$ we see that, to leading order, the
anisotropy $\D$ appears only as a prefactor, i.e.\ as a new temperature
scale. This can be utilized by introducing a rescaled temperature
$\tau = \D T$. Keeping $\a$ and $\tau$ fixed and sending $\D$ to infinity
removes all corrections to ${\cal H}_1$. In this limit the original
XXZ Hamiltonian is replaced by ${\cal H}_1$. The form of the effective
Hamiltonian implies that this should be justified as long as $\D$ is
large enough, say $\D > 10$, since then the second term in (\ref{efham})
becomes small compared to ${\cal H}_1$. At the same time the temperature
$T$ must be small compared to the excitation gap $4J$ of ${\cal H}_I (h_c)$
for the perturbation theory to be applicable.
          
As we have seen, our strong-coupling analysis provides a complete
qualitative picture of the ground state correlation functions in the
region of the phase diagram above the triple point. Clearly, their
values in leading order can be calculated using the exact solution
of $\mcH_1$. Here we will proceed differently. We will take our previous
results for the XXZ chain \cite{TGK09pp}, already conveniently formulated
in terms of integral equations, and consider them in the zero temperature
limit and in the scaling limit. This way we obtain new and explicit
results for the magnetization and some short-range correlation functions.
They are shown in the following sections.

The magnetization and the short-range correlation function in the
full phase diagram of the antiferromagnetic XXZ chain ($\D > - 1$, $h$
and $T$ arbitrary) in the thermodynamic limit can be obtained
by means of the formulae derived in \cite{BDGKSW08,TGK09pp}. This only
requires to solve certain well-behaved linear and non-linear integral
equations numerically and can be done to arbitrary numerical precision.
Examples were worked out in \cite{BDGKSW08,TGK09pp}. In \cite{TGK09pp},
in particular, we noticed the extreme variation of the short-distance
correlation functions close to the triple point. Meanwhile we have worked
this out in more detail.
\begin{figure}[t]
\includegraphics{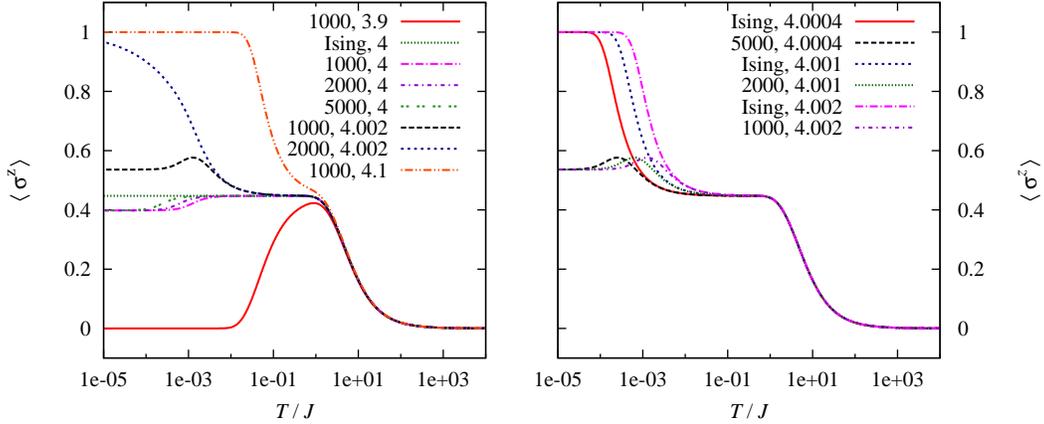}
\caption{(Color online) The one-point function $\left<\sigma^z \right>$
for large values of the anisotropy, fixed $J$ and different values of
$h/J$. The labels in the panels are the tuples $\Delta, h/J$ where `Ising'
denotes the analytic Ising curves.}
\label{fig:ising}
\end{figure}

\begin{figure}[h]
\includegraphics{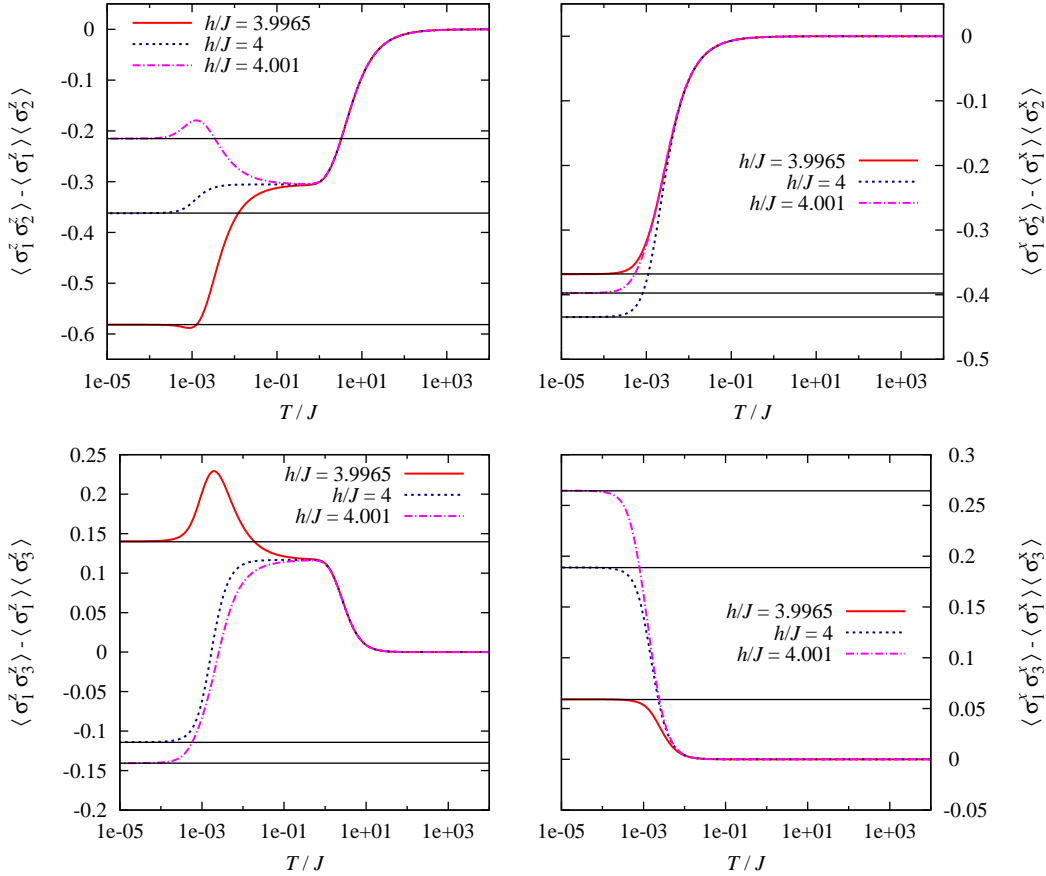}
\caption{(Color online) The temperature dependence of the two-point
functions for two and three neighbouring sites are shown for different
values of the magnetic field for $\Delta=1000$. Additionally, the zero
temperature asymptotic values obtained in \eqref{eq:zerocorr} are shown
as black lines.}
\label{fig:ising_grenzwerte}
\end{figure}

In \cite{TGK09pp} we discussed the temperature behaviour of two-point
correlation functions, e.g.\ of the connected correlation function
$\langle\sigma_1^z\sigma_3^z \rangle - \<\s_1^z\>\< \s_3^z\>$. It turns
out, however, that the signature of the quantum critical point is most
clearly observed in the magnetization per site (which is the only non-%
vanishing one-point function), as it is monotonic in the magnetic field.
We show it in figures~\ref{fig:ising} and \ref{fig:mag3d}. In addition,
the two-point functions for two and three sites are shown in
figure~\ref{fig:ising_grenzwerte}. Note that the non-vanishing of the
transversal correlation functions in the zero temperature limit is due
to the residual quantum mechanical interactions for finite~$\Delta$.

The left panel in figure~\ref{fig:ising} shows how small changes in
the magnetic field or in the coupling $1/\D$ induce large changes in
the low temperature behaviour of the one-point function $\<\s_j^z\>$
due to the proximity of the triple point. The critical cone above the
triple point corresponds to $\a \in [-2, 1]$ (see below). The point
$\D = 1000$, $h = 4.1$ is outside the critical cone on the massive fully
polarized side, while $\D = 1000$, $h = 3.9$ corresponds to a point on
the massive antiferromagnetic side of the phase diagram in
figure~\ref{fig:phasendiagramm_ising}. All other curves in the left panel
of figure~\ref{fig:ising} belong to various values of $\a \in [-2, 1]$.
The curves for $h/J = 4$, i.e.\ for $h = h_c$, in particular, correspond
to $\a = 0$. It can be seen in the figure how, for increasing $\D$, these
curves become more and more similar to the curve for the Ising model at
the critical point, but for small enough temperature always deviate from
the value $\<\s_j^z\> = 1/\sqrt{5}$ of the one-point function of the
Ising chain.

In the right panel of figure~\ref{fig:ising} we compare curves
with different values of the magnetic field and $\a = 0.5$ and curves
for the Ising model with the same values of the magnetic field.
We can identify four different temperature regimes. For very high
temperatures $T \gg h$ the curves are independent of $\Delta$
and $h$. Then, for intermediate temperatures $T \approx h$, they are
independent of the (large) anisotropy and only depend on the magnetic
field. This is where the curves for finite $\Delta$ and the curves of
the Ising model match. We expect that this regime extends down to
$T\approx 10J/\Delta$, where the thermal fluctuations still dominate
the transversal quantum fluctuations. Next comes a regime $T \approx J/\D$,
where thermal fluctuations and transversal quantum fluctuations are of
the same order of magnitude. Here the correlation functions depend on the
anisotropy and the magnetic field. The different curves in this regime
match, if one introduces the rescaled temperature $\tau=T\Delta$, see
section \ref{sec:lowt}. Finally, for very low temperatures $T \ll J/\D$,
the product $\a = \D (h/h_c - 1)$ determines the value of the correlation
functions. This case will be worked out explicitly in the next section.
\vspace{2cm}

\clearpage

\section{Zero temperature limit}
In the following we shall work out explicit formulae for the zero
temperature asymptotics of the magnetization and of a few neighbour
two-point functions on lines of constant $\a = \D (h/h_c - 1)$ in the
vicinity of the triple point. We shall resort to our previous work
\cite{TGK09pp} in which we studied the XXZ chain for $\D > 1$. The simple
idea is to perform the zero temperature limit for large $\D$ and fixed
$\a$ in the special functions of \cite{TGK09pp} that characterize the
correlation functions. For this purpose the form of the non-linear
integral equations as used in \cite{BGKS07} is slightly more convenient
than that of \cite{TGK09pp}, for, in the zero temperature limit, where
the integral equations become linear, the range of integration is over
the Fermi sea and not over its complement. For large $\D$, finally, the
linear integral equations turn into algebraic equations.

Referring to \cite{GKS04a,TGK09pp}, where certain auxiliary functions
$\mfbq$ and $\mfa$ were explained, we introduce the familiar dressed energy
$\ve$ obtained in our setting from $\mfbq$ or $\mfa$ in the zero
temperature limit,
\begin{equation}
\label{eq:dressedenergydef}
     \ve(x) = \lim_{T\rightarrow0}T\ln \mfbq(x)
            = \lim_{T\rightarrow0}-T\ln \mfa(\i x-\eta/2) \epp
\end{equation}
It satisfies the integral equation (see e.g.\ \cite{GKS04a})
\begin{equation}
     \ve(x) = h-4J\sh(\eta)K_{\eta/2}(x)
              -\int_{-\Lambda}^\Lambda \frac{\dx{y}}{\pi} K_\eta(x-y)\ve(y)
\end{equation}
with
\begin{equation}
     K_\eta(x) = \frac{\sh(2\eta)}{2\sin(x+\i \eta)\sin(x-\i \eta)}
\end{equation}
and $\ve(\pm\Lambda)=0$.

Now we use $h=h_c \left(1+\frac{\alpha}{\Delta}\right)$ and expand
the equation up to the order $1/\Delta$. Then
\begin{equation}
     \ve(x) = \frac{h_c \alpha}{\Delta}
            - \frac{h_c \cos(2x)}{\Delta}
	    - \int_{-\Lambda}^\Lambda \frac{\dx{y}}{\pi} \: \ve(y)
\end{equation}
which can be solved explicitly,
\begin{equation}
     \ve(x) = \frac{h_c}{\Delta}
              \left(\frac{\alpha+\sin(\pi c)/\pi}{1+c}-\cos(2x)\right)
\end{equation}
with $c=\frac{2}{\pi}\Lambda$. The Fermi points $\pm\Lambda$ or $\pm c$,
respectively, are determined by the rescaled magnetic field $\alpha$,
\begin{equation}
     \alpha=(1+c)\cos(\pi c)-\frac{1}{\pi}\sin(\pi c) \epp
\end{equation}
Here $c$ is a monotonic function of $\alpha$ for $\alpha\in[-2,1]$. The
boundaries $-2$ and $1$ of the interval correspond to the phase transition
lines (red lines in figure~\ref{fig:phasendiagramm_ising}) determined to
the order $1/\Delta$. For magnetic fields smaller than the lower critical
field $h_e$ or larger than the saturation field $h_s$ the ground state
and, hence, the Fermi points are independent of $h$.

The ground state energy $e$ is given by
\begin{align}
     e = & -\frac{h}{2}+\int_{-\Lambda}^\Lambda
                        \frac{\dx{y}}{\pi}K_{\eta/2}(y)\ve(y)\\
       \approx & - \frac{h_c}{2} \left(1+\frac{\alpha}{\Delta}\right)
                 -\frac{h_c}{\Delta}
		  \frac{\frac{1}{\pi}\sin(\pi c)-\alpha c}{1+c} \epp
\end{align}
Using it one obtains for the magnetization in leading order
\begin{equation}
     m=\frac{\langle\sigma^z\rangle}{2}=\frac{1-c}{2(1+c)} \epp
\end{equation}
Then the leading order magnetic susceptibility is
\begin{equation}
     \chi=\frac{\Delta}{h_c}\frac{1}{(1+c)^3\sin(\pi c)} \epp
\end{equation}
We see that it diverges at the critical point.

In a similar way one can take the zero temperature limit for the functions
that determine the correlation functions, e.g.\ for $\omega$ and $\omega'$
defined in \cite{TGK09pp} and for the auxiliary functions stemming from
linear integral equations (see the appendix, where the zero temperature
limit is obtained from the low temperature approximation discussed in the
next section). This leads to the following expressions for the zero
temperature correlation functions
\begin{subequations}
\label{eq:zerocorr}
\begin{align}
     \left\langle \sigma_1^z\sigma_2^z\right\rangle_{h}
        = & \frac{1-3c}{1+c} \epc \\
     \left\langle \sigma_1^x\sigma_2^x\right\rangle_{h}
        = & -\frac{2\sin(\pi c)}{\pi(1+c)} \epc \\
     \left\langle \sigma_1^z\sigma_3^z\right\rangle_{h}
        = &\frac{1-3c+4c^2}{1+c}-\frac{4\sin^2(\pi c)}{\pi^2(1+c)} \epc \\
     \left\langle \sigma_1^x\sigma_3^x\right\rangle_{h}
        = &\frac{2\sin^2(\pi c)}{\pi^2(1+c)}
	   -\frac{2c\sin(2\pi c)}{\pi(1+c)}+\frac{\sin(2\pi c)}{\pi} \epp
\end{align}
\end{subequations}
In figure~\ref{fig:ising_grenzwerte} we show that the asymptotic
values of the zero-temperature correlation functions match indeed the low
temperature behaviour of the temperature dependent correlation functions
calculated numerically from the integral equations as described in
\cite{TGK09pp}.

\begin{figure}[t]
\includegraphics{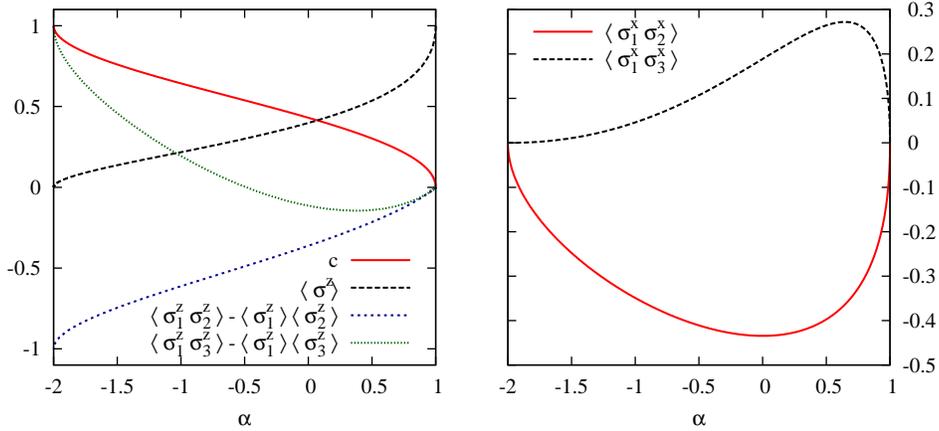}
\caption{(Color online) The asymptotic values of the zero temperature
correlation functions according to equation (\ref{eq:zerocorr}) are shown
as a functions of the rescaled magnetic field $\a$. In addition we show
the normalized Fermi point $c$ as a function of $\alpha$.}
\label{fig:ising_T0}
\end{figure}

In figure~\ref{fig:ising_T0} we show the dependence of the correlation
functions as calculated in (\ref{eq:zerocorr}) on the rescaled magnetic
field $\alpha$. In addition the normalized Fermi point $c$ is shown as
a function of $\alpha$. Both, the Fermi point $c$ and the one-point function
$\langle\sigma^z\rangle$, are monotonic functions of the magnetic field.
The connected two-point function $\left\langle \sigma_1^z\sigma_2^z
\right\rangle-\left\langle \sigma_1^z\right\rangle\left\langle\sigma_2^z
\right\rangle$ is monotonic as well. The other correlation functions
in the figure, however, show non-monotonic behaviour. Both two-point
functions have their extremum above the critical field of the Ising chain,
whereas the minimum of $\left\langle \sigma_1^x\sigma_2^x\right\rangle$
is exactly at the critical field of the Ising chain.

\section{Low temperature behaviour}
\label{sec:lowt}
After the discussion of the zero temperature limit in the previous
section which gave the correct scaling behaviour of the magnetic
field, we would like to discuss small but finite temperatures. For
this purpose we introduce a new rescaled temperature $\tau=\Delta T$
and expand the integral equations to lowest non-vanishing order in
$1/\Delta$ with $\tau$ and $\a$ kept fix. This leads to
\begin{equation}
\label{eq:dressedenergydef_temp}
     \ve(x) = h_c \alpha
            - h_c \cos(2x)
	    + \int_{-\pi/2}^{\pi/2} \frac{\dx{y}}{\pi} \:
              \tau\ln\left(1+\re^{-\ve(y)/\tau}\right)
\end{equation}
where $\ve(x)$ is the finite temperature generalization of
\eqref{eq:dressedenergydef} in the scaling limit. Therefore we use
the same notation for this function. We see that only one interval
$[-\p/2, \p/2]$ of the original contour in the non-linear integral
equation for the full XXZ chain \cite{BGKS07} survives the limit.
The contribution of the remaining part of the contour is exponentially
suppressed for $\Delta\rightarrow\infty$.

The solution of the equation (\ref{eq:dressedenergydef_temp}) is nothing
but the dispersion relation of free fermions on the lattice
\begin{equation}
\ve(x)=\mu-h_c \cos(2x) \epc
\end{equation}
however, with a chemical potential $\mu$ which depends non-trivially on
magnetic field and temperature and which is implicitly determined by the
equation
\begin{equation}
     \mu=h_c \alpha
        + \int_{-\pi/2}^{\pi/2} \frac{\dx{y}}{\pi} \:
          \tau\ln\left(1+\re^{-(\mu-h_c \cos(2y))/\tau}\right) \epp
\end{equation}

The free energy in this approximation is then given by
\begin{equation}
     f= - \frac{h_c}{2} \left(1+\frac{\alpha}{\Delta}\right)
        +\frac {1}{\Delta} \int_{-\pi/2}^{\pi/2} \frac{\dx{y}}{\pi} \:
         \tau\ln\left(1+\re^{-(\mu-h_c \cos(2y))/\tau}\right)
\end{equation}
as for free fermions. Defining
\begin{equation}
     I_n = \int_{-\p/2}^{\p/2} \frac{\rd y}{\p} 
           \frac{\cos(2ny)}{1 + \re^{(\m - h_c \cos(2y))/\tau}} \epp
\end{equation}
the magnetization follows as
\begin{equation}
     m = \frac{1 - I_0}{2(1 + I_0)} \epp
\end{equation}
\begin{figure}[t]
\includegraphics{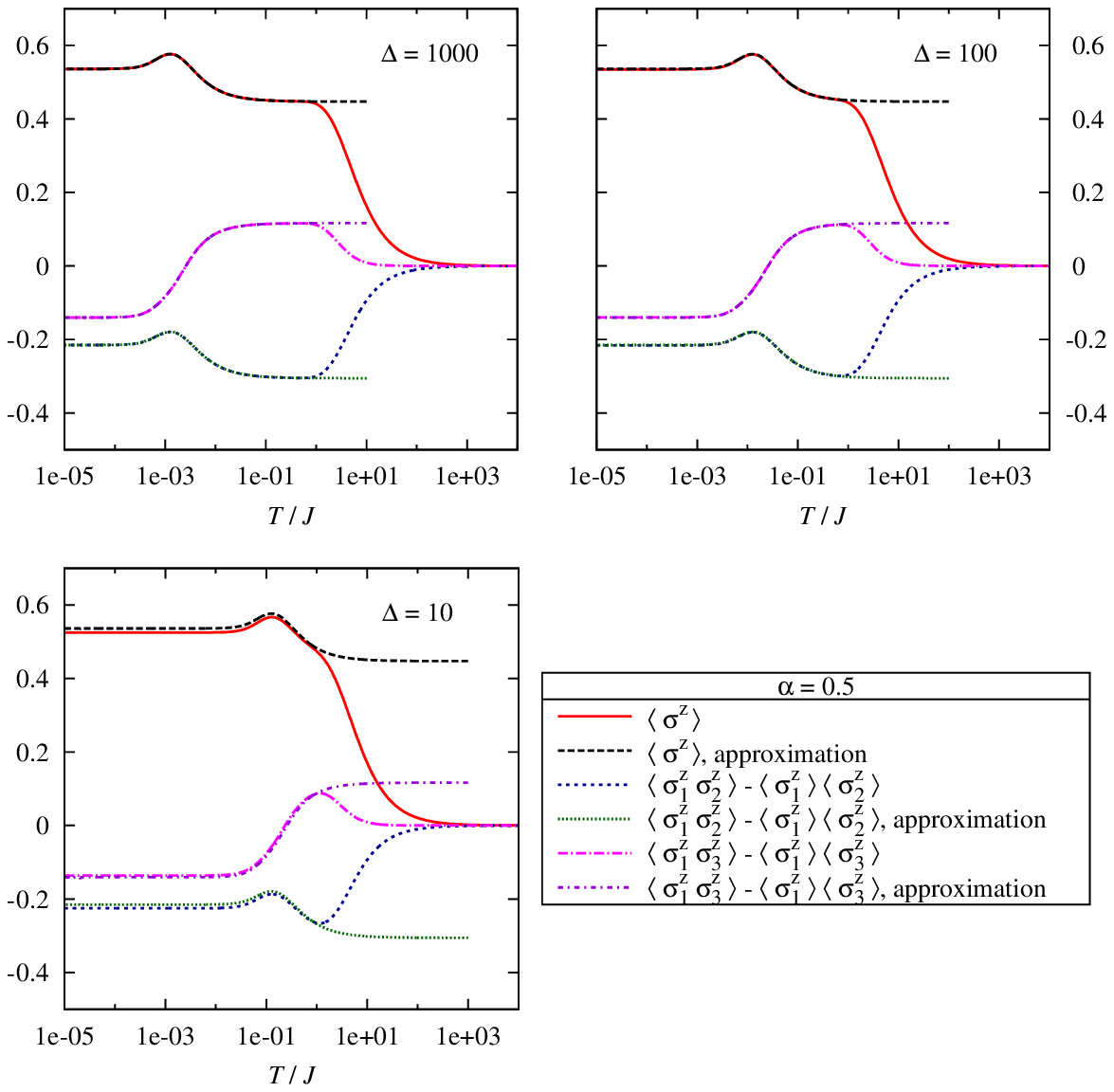}
\caption{(Color online) The longitudinal correlation functions for short
distances as a function of the temperature are shown for different values
of the anisotropy and $\alpha=0.5$.}
\label{fig:ising_05}
\end{figure}

The calculation of the two-point functions within the same approximation
is sketched in the appendix. We obtain
\begin{subequations}
\label{eq:lowtempcorr}
\begin{align}
     \left\langle \sigma_1^z\sigma_2^z\right\rangle_{\tau,h}
        = & \frac{1-3I_0}{1+I_0} \epc \\
     \left\langle \sigma_1^x\sigma_2^x\right\rangle_{\tau,h}
        = & -\frac{2I_1}{1+I_0} \epc \\
     \left\langle \sigma_1^z\sigma_3^z\right\rangle_{\tau,h}
        = &\frac{1-3I_0+4I_0^2}{1+I_0}-\frac{4I_1^2}{1+I_0} \epc \\
     \left\langle \sigma_1^x\sigma_3^x\right\rangle_{\tau,h}
        = &\frac{2I_1^2}{1+I_0}
	   -\frac{4I_0I_2}{1+I_0}+2I_2\epp
\end{align}
\end{subequations}
An important remark is that the structure of the two-point
functions \eqref{eq:lowtempcorr}, unlike the structure of the free energy,
is not the same as for free fermions.
They rather depend on the density-density two-point functions $I_n$ for
free fermions in an interesting way.
\begin{figure}[t]
\includegraphics{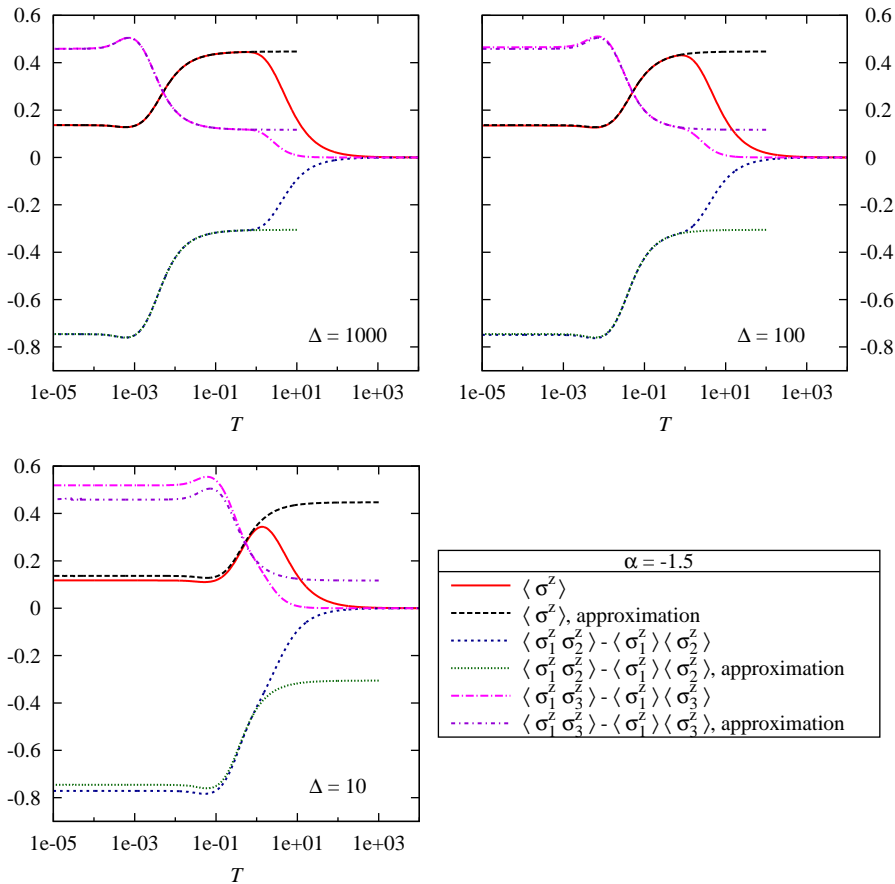}
\caption{(Color online) The longitudinal correlation functions for short
distances as a function of the temperature are shown for different values
of the anisotropy and $\alpha=-1.5$.}
\label{fig:ising_-1}
\end{figure}

In figure~\ref{fig:ising_05} and \ref{fig:ising_-1} the short-range
longitudinal correlation functions are shown for fixed $\alpha$ and
different values of the anisotropy. The exact solutions are compared with
the solutions obtained within the scaling approximation. The agreement
is surprisingly good. The approximate formulae hold up to $T \approx J$.
This is compatible with the fact that the degenerate perturbation
theory should be valid as long as the temperature is small compared to
the excitation gap $4J$. Alternatively we can understand this fact
from the perturbative treatment of the integral equations. As stated
above and explained in some more detail in the appendix, the contribution
of the second contour to the integral equation is exponentially
suppressed with $\Delta$ for small enough $\tau$. On the scale of the
true temperature $T$ this means that it behaves like $\exp(-J/T)$, i.e.,
in particular, in a way independent of $\Delta$.

In order to understand the strange high temperature behaviour of the
correlation functions in the scaling limit, we have to recall that
they are the exact correlation functions of the Hamiltonian ${\cal H}_1$,
which is defined on the restricted Hilbert space with all configurations
containing neighbouring down-spins excluded. In the high temperature
limit, where statistical physics means to simply average over all possible
configurations, the excess of up-spin electrons becomes visible in a
non-vanishing magnetization. We can see in the figure that its high
temperature value becomes independent of the anisotropy, as expected.
Similarly, projecting out configurations with neighbouring down-spins means
that it becomes more likely to have antiparallel neighbours, which causes
the longitudinal neighbour-correlation function to be negative.

The scaling approximation applies to a wide range of the anisotropy.
For $\Delta=100$ and $\Delta=1000$ the deviation from the exact solution is
hardly recognizable, and the approximation is still in good qualitative
agreement with the exact solution for $\Delta=10$. Comparing the cases
$\alpha=0.5$ in figure~\ref{fig:ising_05} and $\alpha=-1.5$ in figure~%
\ref{fig:ising_-1} one sees that for $\Delta=10$ the offset in the zero
temperature limit is larger in the latter case. This is in agreement with
the observation that the critical lines in the zero-temperature phase
diagram in figure~\ref{fig:phasendiagramm_ising} relate differently
to $\a$. The phase transition line to the fully polarized state belongs
exactly to $\a = 1$, while the line marking the transition to the N\'eel
phase belongs to $\a = - 2$ only asymptotically and deviations become
visible for $\D \approx 10$. For this reason we generally expect larger
deviations for smaller $\a \in [-2,1]$.

In general both figures show that in the regime $T \approx J/\Delta$,
in which the correlation functions depend on the anisotropy and the
magnetic field, the dependence on the anisotropy can be removed by
introducing the rescaled temperature $\tau$. Since the scaling
approximation is valid for temperatures up to $T \approx J$ and since
the XXZ chain is well approximated by the Ising chain for temperatures
higher then $T \approx 10J/\Delta$, the critical regime of the XXZ
chain above the triple point is fully accessible by simple approximations
(either Ising or ${\cal H}_1$) as long as $\D > 10$. Outside the critical
cone the Ising approximation works well for $\D > 10$ and all temperatures
anyway.

\section{Conclusion}
Based on the exact solution for the thermodynamics and short-%
distance correlation functions \cite{TGK09pp} we have examined the
vicinity of the triple point of the XXZ chain in the critical regime.
The low temperature physics in this region of the phase diagram and
the nature of the singularity at the critical point can be understood
qualitatively be means of degenerate perturbation theory around the
ground state of the critical Hamiltonian ${\cal H}_I (h_c)$. From the
form of the second order effective Hamiltonian (\ref{efham}) it follows
that a small scale $T \approx J/\D$ exists on which the correlation
functions show strong variations and that below this scale the correlation
functions must be constant on straight lines of fixed $\a =
\D (h/h_c - 1)$. Using the exact solution, formulated in terms of certain
linear and non-linear integral equations \cite{TGK09pp}, we have obtained
a full quantitative understanding of the vicinity of the critical point
at any temperature. We have seen that, as long as $\D > 10$ the
full XXZ Hamiltonian in the critical phase above the triple point
is well approximated by the effective Hamiltonian ${\cal H}_1$ as
long as $T < J$ and by $H_I (h_c)$ for $T > 10J/\D$. In particular,
for the interesting intermediate and low temperature regime $T < J$
it is enough to consider ${\cal H}_1$. This is interesting, since
${\cal H}_1$ belongs to the relatively simple class of Hamiltonians
similar to the impenetrable Bose gas \cite{AlBa99b}. The Bethe ansatz
solution of models in this class is simple enough to admit for the
derivation of finite temperature dynamical correlation functions in a
closed form called determinant representation \cite{KBIBo}. A determinant
representation for ${\cal H}_1$ was derived in \cite{AbPr02}. It could
be the future starting point for an analysis of the asymptotic behaviour
of the dynamical correlation functions in temperature regions in which
conformal field theory is no longer a valid approximation. Such an analysis
would be complementary to the perturbative study of the finite
temperature dynamical correlation functions in the massive N\'eel ordered
regime recently presented in \cite{JGE09}.

\section*{Acknowledgment}
The authors are grateful to C. Malyshev for bringing reference
\cite{AlBa99b} to their attention and to A. Pronko for providing a copy
of his paper \cite{AbPr02}. CT likes to acknowledge support by
the Volkswagen Foundation.

\appendix

\section*{Appendix: Low temperature approximation and zero temperature
limit of the linear integral equations}
We are referring to the notation and definitions of \cite{TGK09pp}.
In order to calculate the low temperature limit of the two-point functions
we have to begin with the functions $G$ and $G'$. For these functions
a shift as in \eqref{eq:dressedenergydef} has to be made. We define 
\begin{equation} \label{eq:gprimedef}
     g^-_\nu(x):=-G(\i x-\eta/2,\i \nu) \epc \qd
    {g'}^-_\nu(x):=-\frac{G'(\i x-\eta/2,\i \nu)}{\eta}
\end{equation}
which is consistent with the notation used in \cite{TGK09pp} except for
the denominator $\eta$ in the second equation introduced for convenience
here.

In the low temperature limit for large $\Delta$ the linear integral
equations for $G$ and $G'$ imply
\begin{subequations}
\begin{align}
     g_\nu^-(x) & = -2K_{\eta/2}(x-\nu)
                  -\int_{-\pi/2}^{\pi/2} \frac{\dx{y}}{\pi} K_\eta(x-y)
		   \frac{g_\nu^-(y)}{1+\re^{(\mu-h_c\cos(2y))/\tau}}
		   \epc \\[1ex]
     {g'}_\nu^-(x) & = -\i \cot(x-\nu- \frac{\i \eta}{2})
       -\int_{-\pi/2}^{\pi/2} \frac{\dx{y}}{\pi} L_\eta(x-y)
        \frac{g_\nu^-(y)}{1+\re^{(\mu-h_c\cos(2y))/\tau}} \notag \\
	& \mspace{216.mu}
       -\int_{-\pi/2}^{\pi/2} \frac{\dx{y}}{\pi} K_\eta(x-y)
        \frac{{g'}_\nu^-(y)}{1+\re^{(\mu-h_c\cos(2y))/\tau}} \epc
\end{align}
\end{subequations}
where
\begin{equation}
     L_\eta(x) = \frac{\i \sin(2x)}{2\sin(x+\i \eta)\sin(x-\i \eta)} \epp
\end{equation}
The second part of the integration contour vanishes in this limit as it
gives only exponentially small contributions for $\Delta\rightarrow\infty$,
similar to \eqref{eq:dressedenergydef_temp}.

We solve these equations up to the order $\exp(-2\eta)$ and obtain
\begin{subequations}
\label{gandgpexp}
\begin{align} 
     & g_{\nu}^-(x) = -\frac{2}{1+I_0}
        +\re^{-\eta}\left[\frac{4I_1}{1+I_0}\cos(2\nu)
	                -4\cos(2(x-\nu))\right] \notag \\
     & \mspace{48.mu}
       +\re^{-2\eta} \left[\frac{4I_2}{1+I_0}\cos(4\nu)
                        -\frac{4I_1^2}{(1+I_0)^2}
			-4\cos(4(x-\nu))
			+\frac{4I_1}{1+I_0}\cos(2x)\right]
			\epc \displaybreak[0] \\[2ex]
     & {g'}_{\nu}^-(x) =
        \frac{1}{1+I_0}+\re^{-\eta}\left[2\re^{-\i 2(x-\nu)}
	                 -\frac{2I_1}{1+I_0}\re^{\i 2\nu}\right]
        +\re^{-2\eta}\biggl[ 2\re^{-\i 4(x-\nu)} \notag \\
     & \mspace{135.mu} \left.+\frac{\i 4I_1}{1+I_0}\sin(2x)
        -\frac{2I_1}{1+I_0}\cos(2x)
	-\frac{2I_2}{1+I_0}\re^{\i 4\nu}
	+\frac{2I_1^2}{(1+I_0)^2}\right] \epp
\end{align}
\end{subequations}
Finally we need the low temperature limit of $\omega$ and $\omega'$
defined in \cite{TGK09pp}. One easily finds that
\begin{subequations}
\begin{align}
     & \omega(\nu_1,\nu_2) = K_\eta(\tilde{\nu}_1-\tilde{\nu}_2)
        + 2 \int_{-\pi/2}^{\pi/2} \frac{\dx{x}}{\pi}
	     K_{\eta/2}(x-\tilde{\nu}_2)\frac{g_{\tilde{\nu}_1}^-(x)}{1+\re^{(\mu-h_c \cos(2x))/\tau}} \epc \\[1ex]
     & \frac{\omega'(\nu_1,\nu_2)}{\eta}
        =  -L_\eta(\tilde{\nu}_1-\tilde{\nu}_2)
	  -\int_{-\pi/2}^{\pi/2} \frac{\dx{x}}{\pi}
	     \i \cot(x-\tilde{\nu}_2-\i \eta/2)\frac{g_{\tilde{\nu}_1}^-(x)}{1+\re^{(\mu-h_c \cos(2x))/\tau}}
	     \notag \\[-1ex] & \mspace{270.mu}
	  +2\int_{-\pi/2}^{\pi/2} \frac{\dx{x}}{\pi}
	     K_{\eta/2}(x-\tilde{\nu}_2)\frac{{g'}_{\tilde{\nu}_1}^-(x)}{1+\re^{(\mu-h_c \cos(2x))/\tau}} \epc
\end{align}
\end{subequations}
where we used the notation $\tilde{\nu}_j=-\i \nu_j$. Inserting
(\ref{gandgpexp}) here and using equations (24) and (25) of \cite{TGK09pp}
we obtain the equations (\ref{eq:lowtempcorr}).

For the zero temperature limit integrals over the Fermi weight of the
form 
\begin{equation}
     \int_{-\pi/2}^{\pi/2} \frac{\dx{x}}{\pi} \:
        \frac{f(x)}{1+\re^{(\mu-h_c \cos(2x))/\tau}}
\end{equation}
have to be changed into integrals over the Fermi sea
\begin{equation}
\int_{-\Lambda}^\Lambda \frac{\dx{x}}{\pi} \: f(x) \epp 
\end{equation}
The functions $I_n$, for instance, that determine the correlation functions
in the low temperature scaling approximation have the following zero
temperature limit
\begin{subequations}
\begin{align}
     I_0 &\rightarrow c \epc\\
     I_1 &\rightarrow \frac{\sin(\pi c)}{\pi} \epc\\
     I_2 &\rightarrow \frac{\sin(2\pi c)}{2\pi} \epp
\end{align}
\end{subequations}
With this limit introduced into \eqref{eq:lowtempcorr} one obtains
the zero temperature correlation functions \eqref{eq:zerocorr}.


\providecommand{\bysame}{\leavevmode\hbox to3em{\hrulefill}\thinspace}
\providecommand{\MR}{\relax\ifhmode\unskip\space\fi MR }
\providecommand{\MRhref}[2]{%
  \href{http://www.ams.org/mathscinet-getitem?mr=#1}{#2}
}
\providecommand{\href}[2]{#2}

\end{document}